\documentstyle[aps,prb,multicol,epsf,picinpar]{revtex} 

\def\ep {\epsilon}

\def\e2 {\epsilon-\epsilon_k}
\def\be {\begin{equation}}
\def\ee {\end{equation}}
\def\bea {\begin{eqnarray}}
\def\eea {\end{eqnarray}}
\def\om {\omega}

\begin{document}
\draft           
\draft
\title{Low temperature dephasing saturation from elastic magnetic
spin disorder and interactions }

\author{ George Kastrinakis}

\address{Institute of Electronic Structure and Laser (IESL), Foundation for
Research and Technology - Hellas (FORTH), 
P.O. Box 1527, Iraklio, Crete 71110, Greece$^*$}

\date{April 26, 2005}

\maketitle

\begin{abstract} 

We treat the question of the low temperature behavior of 
the dephasing rate of the electrons in the presence 
of elastic spin disorder scattering and interactions. 
In the frame of a self-consistent diagrammatic treatment, 
we obtain saturation of the dephasing rate in the limit of low temperature
for magnetic scattering, in agreement with the non-interacting case.
The magnitude of the dephasing rate is set by the strength of
the magnetic scattering rate. 
We discuss the agreement of our results with relevant experiments.
\end{abstract}

\vspace{.3cm}

An important quantity in disordered electronic systems is 
the dephasing rate $\tau_\phi^{-1}$. It provides a measure of the loss of 
coherence of the carriers, but in the two-particle channel - c.f. eq. (1)
below. Decoherence arises from coulombic interactions,
scattering by phonons, magnetic fluctuations etc. The saturation of the 
dephasing rate at low temperature $T$ seen in {\em numerous} 
experiments \cite{lib,mo1,mo2,bir,iko,nat,mar,piv,gou,lin,lin2,pie,pie2} 
has attracted a vigorous interest, especially given the longstanding 
theoretical prediction for a vanishing $\tau_{\phi}^{-1}$ 
as the temperature $T \rightarrow 0$.\cite{aak,fa,aag,aav,rsc,vk1,vk2,vde}.

Previous theoretical studies 
\cite{aak,fa,aag,aav,rsc,vk1,vk2,vde,gz1,gzs,ifs,zvr} 
have focused on the calculation
of $\tau_{\phi}^{-1}$ in the absence of spin-scattering disorder.
The majority of these studies predict, correctly, a vanishing 
$\tau_\phi^{-1}(T\rightarrow 0)$ .
Here we determine and calculate the factors which contribute to dephasing
in the presence of spin-scattering disorder. The saturation obtained
allows for the consistent elucidation of this puzzle. 

In the presence of spin-less disorder, the cooperon 
(particle-particle diffusion correlator - c.f. fig. 1) is given by
\be
C_0(q,\om) = \frac{1}{2\pi N_F \tau^2}
\frac{1}{Dq^2 - i \om + \tau_{\phi}^{-1}} \;\;.
\ee
$D$ is the diffusion coefficient,
$N_F$ is the density of states at the Fermi level and $\tau^{-1}$ 
the total impurity scattering rate. We work in the diffusive regime
$\epsilon_F \tau >1$ ($\hbar=1$), $\epsilon_F$ being the Fermi energy.

With spin-disorder present, the cooperon 
becomes spin-dependent. The relevant terms $C_i$ are shown in fig. 1.
We start by giving the explicit form of these $C_{0,1,2}^o$ {\em without} 
a dephasing rate. $C_0^o$ acquires a {\em finite spin-dependent term} in
the denominator, which is crucial for the determination of the dephasing
rate - c.f. below.

For the case $, \tau_S^{-1}>0, \tau_{so}^{-1}=0$ - 
with $\tau_S^{-1}$ the magnetic impurity
scattering rate  and $\tau_{so}^{-1}$ the
spin-orbit impurity scattering rate - the cooperons are given by 
\bea
C_0^o(q,\om) = \frac{1}{2\pi N_F \tau^2} \;  \label{coos}
\frac{1}{Dq^2 - i \om + 2/(3 \tau_{S})} \;\;,  \\
C_{1,2}^o(q,\om) = b_{1,2} \Big\{ \frac{1}{Dq^2 -i \om + 2/(3 \tau_S)}
-\frac{1}{Dq^2 -i \om + 2 / \tau_S } \Big\} \;\;, \nonumber
\eea
with $b_1=(3\tau_S/(2 \tau) -2)/(2u), b_2=1/u, u=4\pi N_F \tau^2 $.
Based on eqs. (\ref{coos}), we expect a saturation of the dephasing rate.
The simple diffusion pole is "cut-off" by the constant terms proportional 
to $\tau_S^{-1}$. On symmetry grounds, the spin-conserving Coulomb
interaction cannot eliminate these terms.

We emphasize that the impurity scattering considered is elastic, bulk-type.
Interfacial impurity scattering, though similar to bulk-type, is expected 
to differ in detail.

To calculate the dephasing rate, we write down and 
solve the appropriate coupled equations for {\em all} three renormalized
cooperons $C_i(q,\om)$, $i=0,1,2$. We note that usually the terms containing 
the factors
$d_i$ and $h_i$ below are completely omitted. The equations are shown 
schematically in fig. 2:
\bea
C_0 = C_0^o + C_0^o Y_0 C_0 \;\;,  \label{exsc1}\\
C_1 = C_1^o + C_1^o W_a + C_2^o W_b \;\;, \label{eco1} \\
C_2 = C_2^o + C_1^o W_b + C_2^o W_a \;\; \label{eco2} ,
\eea
with 
\be
W_a = Y_n C_1 + Y_r C_2\;,\; W_b = Y_n C_2 + Y_r C_1 \;\; ,
\ee
and
\be
Y_0 = (1+h_0) \Sigma_0 + d_0 \Sigma_1 \;,\;
Y_n = (1+h_1) \Sigma_1 + d_1 \Sigma_0 \;, \; 
Y_r = (1+h_2) \Sigma_2 + d_2 \Sigma_0  \;.
\ee
In fig. 3 we show explicitly the components of the self-energy terms $Y$
- c.f. the figure legend for further details.
Here, $h_0 = h_1 = h_2 = -2/(\pi \ep_F\tau)$, 
$d_0=d_1= \{1/(\ep_F\tau) + 4\pi\} \{\tau_{so}^{-1}-\tau_S^{-1}\}
/(2\pi^2\ep_F )$ and 
$d_2=-2 \{1 + 2 \pi \ep_F\tau \}\{\tau_{so}^{-1}-\tau_S^{-1}\}
/(\pi \ep_F )$.
The terms containing the factors $d_i$ - with a spin impurity line either
looping
around the cooperon or crossing it - provide the coupling between
the spin-independent and the spin-dependent cooperons, as they produce
{\em spin flipping} - c.f. the spin configuration of $C_2$ in fig. 1 and
the figure legend.
Also, note the minus relative sign 
between $\tau_{so}^{-1}$ and
$\tau_S^{-1}$ in $d_i$, coming from the respective spin flipping disorder
vertices \cite{berg}. 
This has the peculiar effect that for $\tau_{so} = \tau_S$ 
the spin disorder signature {\em disappears} in $\tau_{\phi o}$ - c.f.
eq. (\ref{ext}) below.

In the spirit of ref. \cite{fa} we obtain for the basic components $\Sigma_i$
of the self-energy
\be
\Sigma_i \equiv \Sigma_i(q=0,\omega=0)\simeq \\
- c_o \; \sum_q \int_{-\infty}^{\infty} d\om'
\frac{ C_i(q,\om'+i 0) \; \text {Im} V(q,\om'+i 0)}{\sinh(\om'/T)} \;\;,
\label{es0}
\ee
where
\be
V(q,\om)=\frac{v_q}{1+v_q \Pi(q,\om)} \;\;, \;\; \\
\Pi(q,\om) = \frac{N_F Dq^2}{Dq^2 - i\om} \;\;. \label{pot}
\ee
Here $v_q$ is the bare Coulomb interaction and 
$c_o=8 N_F^2 \tau^4 $. 

In the foregoing we make the approximation \cite{fa} 
\be
\int_{-\infty}^{\infty} d\om \frac{F(\om)}{\sinh(\om/T)}
\simeq T\int_{-T}^T d\om \frac{F(\om)}{\om} \;\;. \label{ese}
\ee
The bosonic modes with energy greater than $T$ manifestly do {\em not}
contribute to the self-energy and the dephasing process, as also emphasized 
in refs. \cite{aav,vde}.

In the limit $\ep_F \tau\gg 1$, we can decouple a
$2\times 2$ system of equations involving only $C_{1,2}$, to facilitate the
solution of eqs. (\ref{exsc1}-\ref{eco2}).
Then eq. (\ref{exsc1}) yields directly for the dephasing rate 
$\tau_{\phi o}^{-1}$ - note the index $o$ - in the 
denominator of $C_0$

\be
\tau_{\phi o}^{-1}(T) = - \frac{\Sigma_0+d_0 \Sigma_1}
{2 \pi N_F \tau^2} \;\;. 
\label{ext}
\ee
This equation means that the
dephasing in the spin-dependent cooperon channels also contributes 
to the dephasing rate in the spin-independent channel.
We note that in the limit  $\ep_F \tau\gg 1$ the factor $d_0$ in
(\ref{ext}) is the only one remaining among the $d_{0,1,2}$ and $h_{0,1,2}$
appearing in eqs. (\ref{exsc1})-(\ref{eco2}). But as shown below, the
factor $d_0 \Sigma_1$ is too small in comparison to the factors 
$\tau_{S}^{-1}, \tau_{so}^{-1}$ in the denominator of $C_0$.

We look in detail at the case of pure magnetic scattering, 
i.e. $\tau_S^{-1}>0$ and $\tau_{so}^{-1}=0$. In Appendix B we discuss
the case of pure spin-orbit scattering.

Diagonalizing the system formed by eqs. (\ref{eco1},\ref{eco2}) yields for 
the cooperons $C_{1,2}$ 
\be
C_{1,2}(q,\om) = S_{1,2} \Big[ \frac{1}{Dq^2 -i \om + R_-} - 
\frac{1}{Dq^2 -i \om + R_+} \Big] \;\;. \label{cosog}
\ee
Here $M_1= [b_1 m_1 m_2 + (b_2^2-b_1^2) m_0 \Sigma_1]/m, 
M_2 =[b_2 m_1 m_2 - (b_2^2-b_1^2) m_0\Sigma_2]/m, 
R_{\pm}= [m_1 m_2\pm m_0 (b_2\mp b_1)(\Sigma_1 \mp \Sigma_2)]/m,
X= 2(b_2 \Sigma_1+b_1 \Sigma_2)$, $m_1 = 2/(3 \tau_S), 
m_2=2/\tau_S, m_0=m_2-m_1, m=m_1+m_2$.
The derivation of eqs. (\ref{cosog}) can be found in Appendix A.
We note that it is important to keep the term $b_j (Dq^2 -i \om)$ in
$S_j$, to arrive at the correct solution. Further, eqs. (\ref{cosog})
are valid in any dimension.

In 2-D, $v_q=2\pi e^2/q$, and, following \cite{fa,rsc}, we take 
$V(q,\om)=\frac{2 \pi e^2}{q} \frac{D q^2-i\om}{D\kappa q-i\om}$,
with $\kappa = 4 \pi N_F e^2$.

To calculate the self-energies, we first evaluate the integral in eq. 
(\ref{ese})
\be
I_j(q) = \int_{-T}^T d\om \;
\frac{b_j (Dq^2-i\om)+M_j}{(D\kappa q)^2+\om^2} \;\Big\{
\frac{1}{D q^2-i\om+R_-} - \frac{1}{D q^2-i\om+R_+} \Big\}\;\;.
\ee
Since we are interested in the low $T$ limit, we take
\be
D q^2 > T \;\;, \label{minq}
\ee
obtaining
\be
I_j(q) \simeq 2 T \frac{b_j Dq^2+M_j}{(D\kappa q)^2} \;\Big\{
\frac{1}{D q^2+R_-} - \frac{1}{D q^2+R_+} \Big\} 
\;\;.
\ee
Subsequently, we evaluate
\be
\int_{\sqrt{T/D}}^{\sqrt{z/D}} I_j(q) \; (\kappa-q)q \;dq \;\;,
\ee
with
\be
z= \text{max} \; Dq^2 = \tau^{-1} \;\;.  
\ee
Finally we obtain the following equations for the self-energies 
\bea
\Sigma_j = \frac{T^2 a_{2D}}{X} 
\Big\{ 
\frac{M_j-b_j R_+}{R_+} \Big[ \kappa \ln\Big( \frac{R_++z}{R_++T}\Big)
- 2 \sqrt{\frac{R_+}{D} }\Big( \arctan \sqrt{\frac{z}{R_+}} - 
\arctan \sqrt{\frac{T}{R_+} } \Big) \Big]    \label{exsi2} \\ 
- \frac{M_j-b_j R_-}{R_-} \Big[ \kappa \ln\Big( \frac{R_-+z}{R_-+T}\Big)
- 2 \sqrt{\frac{R_-}{D} }\Big( \arctan \sqrt{\frac{z}{R_-}} - 
\arctan \sqrt{\frac{T}{R_-} } \Big) \Big] 
+\kappa \; M_j \Big(\frac{1}{R_-}-\frac{1}{R_+}\Big)\ln\Big(\frac{z}{T}\Big)
\Big\} \;\;, \nonumber
\eea
with $j=1,2$
and $a_{2D}=2 c_o e^2/(D \kappa^2)$. 

The solution is 
\be
\Sigma_j = s_j T^2 \;\;, \;\; 
s_j = b_j a_{2D} \tau_S \left\{ \sqrt{\frac{2}{D \tau_S}}
\arctan \sqrt{2 z \tau_S} - \kappa \ln\left( 1+ 2 z \tau_S \right) \right\}
\;\; ,
\ee
which is valid for 
\be
\Sigma_j  \ll  1/(2 \tau_S) \;\;.
\ee
This condition accompanies the solutions for $\Sigma_j$ in 1-D and 
3-D as well.

We consider the so-called weak localization contribution to the
conductivity, given by the sum of  \cite{berg}
\be
\delta \sigma_o = -\frac{ e^2 D u}{\pi} 
\sum_q \big\{ C_0(q,0)+C_2(q,0) \big\} \label{swl2}
\;\;, 
\ee
which in a magnetic field $H$ perpendicular to the 2-D system becomes
\bea
\delta \sigma_o = -\frac{ e^2 D u}{\pi}
\frac{e H}{\pi} \sum_{n=0}^{N_H} \Big\{
\frac{1}{2\pi N_F \tau^2} \; \frac{1}{4 D e H(n+1/2) + 2/(3\tau_{S})+ 
\tau_{\phi o}^{-1} }   \label{swl2sh}  		\\    \nonumber
+ \frac{4 b_2 e D H(n+1/2)+ M_2}{X} \Big( \frac{1}{4 D e H(n+1/2) + R_-}
-\frac{1}{4 D e H(n+1/2) + R_+} \Big)  \Big\}  \;\;,
\eea
with $N_H=1/(4D e \tau H)$.

If this formula were 
fit to the 2-D formula without magnetic (or spin-orbit) scattering
\be
\delta \sigma_o = -\frac{ 2 e^3 D H}{\pi^2} \sum_{n=0}^{N_H} 
\frac{1}{4 D e H(n+1/2) + 
\tau_{\phi}^{-1} }    	\;\;.  \label{2dwl}
\ee
saturation of the dephasing is obtained due (mostly) to the factor 
$2/(3\tau_{S})$ in the denominator of $C_0$. The factors $R_\pm$ in eq. 
(\ref{swl2sh}) 
satisfy $R_\pm = 1/(2 \tau_S) + O(T^2)$. As a result, the contribution of
the $C_2$ term is small, because $R_+ - R_- = O(T^2)$.
The same applies to 1-D and 3-D, with the power law being $T^{3/2}$ in 1-D,
as shown below.

In 1-D, $v_q = 2  e^2 \ln(q_m/q)$, $q_m$ being the inverse of the largest 
transverse dimension (width) of the system. Here, 
$Im V(q,\om)=-\frac{4e^4 N_F \om \; Dq^2 \;
\ln^2(q_m/q)}{\om^2+(Dq^2)^2 [1+2 e^2 N_F \ln(q_m/q)]^2}$.

To calculate the self-energies in eq. (\ref{es0}), we first evaluate the 
integrals
\bea
L_j(\om) = 
\int_{0}^{z} dx \frac{\sqrt{x} \; (b_j(x- i\om)+M_j)}
 { B^2 x^2+\om^2}
\Big\{\frac{1}{x-i\om+R_-}  -  \frac{1}{x-i\om+R_+} \Big\}  \label{ol1}
\\ \nonumber
=-2\frac{(M_j-b_j R_-)\sqrt{R_- -i\om} 
\; \text{arctan}{\sqrt{\frac{z}{R_--i\om}}} } 
{ \; \{ B^2(R_--i\om)^2+\om^2\} } \\ \nonumber
+2\frac{ (M_j-b_j R_+)\sqrt{R_+-i \om} 
\; \text{arctan}{\sqrt{\frac{z}{R_+-i\om}}} } 
{ \; \{ B^2(R_+-i\om)^2+\om^2\} } \\ \nonumber
+ \frac{ \{i M_j B+ b_j \om(B-1)\} 
\arctan{\sqrt{\frac{i Bz}{\om} }}}{ B^{3/2} \sqrt{-i \om} } \; \left\{
\frac{1}{ iB R_- +\om(B-1) }- \frac{1}{ iB R_++\om(B-1) } \right\} \\ \nonumber
+ \frac{\{i M_j B+ b_j \om(B+1)\} 
\arctan{\sqrt{\frac{-i Bz}{\om} }}}{ B^{3/2} \sqrt{i \om} } \; \left\{
\frac{1}{ iB R_-+\om(B+1) }- \frac{1}{ iB R_++\om(B+1) } \right\} 
\;\;. 
\eea
Here $x=D q^2$, $z = \text{max} \; x = \tau^{-1}$ as before, and we 
approximated the term $\ln(q_m/q)$ by its average, taking
\be
B = 1+e^2 N_F <\ln(q_m^2/q^2)> \;\;\; > 1 \;\;. \label{avg}
\ee

Then, we consider the low $T$ limit $z\gg T > |\om|$, obtaining
\be
L_j(\om) =  \frac{M_j \pi }{ \sqrt{2 \om B^3 } } 
\left\{ \frac{1}{R_-}-\frac{1}{R_+} \right\} + \text{const.} \;\;,
\ee
and we evaluate 
\be
\int_{-T}^T\; L_j(\om)\; d\om  \;\;.
\ee

Thus we obtain the self-energy equations 
\bea
\Sigma_j = \frac{T^{3/2} a_{1D} M_j }{X } \left\{\frac{1}{R_-}-\frac{1}{R_+} 
\right\}  \;\;, \label{eq1d} 
\eea
with $j=1,2$, and $a_{1D}= \sqrt{2} c_o (B-1)^2/(N_F B^{3/2}) $.
The solution is
\be
\Sigma_j = s_j \; T^{3/2} \;\;, \;\; 
s_j=\frac{\sqrt{2}\; b_j \tau_S c_o(B-1)^2} { N_F B^{3/2} } \;\;.
\ee

In 3-D we take $V(q,\om)=\frac{4\pi e^2}{q^2} \frac{Dq^2-i \om}{P -i \om}$, 
with $P=4\pi e^2 N_F D$. To calculate the self-energies of eq. (\ref{es0}), 
we first evaluate the integrals
\bea
K_j(q) = \int_{-T}^T \; d\om \;
\; \frac{ b_j(D q^2 -i\om)+M_j }{P^2+\om^2}
\Big\{\frac{1}{D q^2-i\om+R_-}  -  \frac{1}{D q^2-i\om+R_+} \Big\}  \\
\simeq 2 T \frac{ b_j D q^2 +M_j }{P^2}
\Big\{\frac{1}{D q^2+R_-}  -  \frac{1}{D q^2+R_+} \Big\} 
\;\;. \nonumber
\eea
Then, evaluating
\be
\int_{\sqrt{T/D}}^{\sqrt{z/D}} K_j(q) \; \big\{ Dq^2-P\big\} \;dq \;\;,
\ee
yields the self-energy equations
\bea
\Sigma_j = \frac{T^2 a_{3D}}{X} \Big\{ 
\frac{(M_j-b_j R_+)(P+R_+)}{\sqrt{D R_+}}\arctan \sqrt{\frac{z}{R_+} }
-\frac{(M_j-b_j R_-)(P+R_-)}{\sqrt{D R_-}}\arctan \sqrt{\frac{z}{R_-} }  
\label{es3d} \\
+b_j (R_+-R_-) \sqrt{\frac{z}{D}} \Big\} \;\;. \nonumber
\eea
Here $a_{3D}=4 c_o e^2/(\pi P^2 )$. We obtain the solution
\be
\Sigma_j= s_j T^2 \;\;,\;\; s_j = \frac{b_j a_{3D}}{2 \sqrt{D}}
\left\{ \sqrt{z} - \sqrt{2 \tau_S} 
\left(\frac{1}{2 \tau_S}+P \right)\arctan \sqrt{2\tau_S z}  \right\} \;\;.
\ee

From the above we see that
$\tau_{\phi o}^{-1}(T \rightarrow 0)$ 
is much smaller than both $\tau^{-1}$ and 
$\tau_{sp}^{-1}=\tau_{so}^{-1}+\tau_S^{-1}$.
Actually, $\tau_{sp}^{-1}$ dominates over $\tau_{\phi o}^{-1}(T\rightarrow 0)$
in the denominator of $C_0$, causing a 'saturation' of this 
contribution. 

Besides the spin-scattering rate terms appearing in the denominator of
$C_0$, the dephasing rate probed in experiments is also set by the
terms $R_{\pm}$ - c.f. eq. (\ref{swl2sh}), pure magnetic scattering 
case. As mentioned in Appendix B, it is expected that in the presence of the 
Coulomb interaction
the simple diffusion pole in $C_2$ - c.f. eq. (\ref{cooso}) - survives intact,
in the pure spin-orbit scattering case, thus 
yielding absence of dephasing saturation. 

We note that the idea
that magnetic scattering may cause saturation has been recently suggested
in ref. \cite{pie2}.
The situation here is to be contrasted with the absence of spin disorder,
where it has been shown that $\tau_{\phi}^{-1}(T\rightarrow 0) \rightarrow 0$,
e.g. in 2-D $\tau_{\phi}^{-1}(T\rightarrow 0) \propto T 
\rightarrow 0$ \cite{aak,fa,aag,aav,rsc,vk1,vk2,vde}.
Moreover, we should point out that other self-energy processes, 
which are first order in the interaction $V(q,\omega)$, e.g. with $V$ 
crossing diagonally the cooperon - c.f. e.g. \cite{aav}, do not modify 
qualitatively these results.

Now, the total correction to the conductivity can be written as
\be
\delta \sigma_{tot} = \delta \sigma_o + \delta \sigma_{I} \;\;, \label{stot}
\ee
where the first term is the cooperon contribution of eq. (\ref{swl2})  
and the second term due to interactions, involving additional and 
more complicate terms. If $ \delta \sigma_{I}(T\rightarrow 0)$ contains
non-saturating terms, then the picture so far presented should change.

The majority of experiments 
show saturation of $\tau_\phi^{-1}$ in the low temperature limit.
The samples in which saturation is observed probably contain magnetic 
impurities, even in minute quantities.
Also, some of the samples, in which  
dephasing saturation is observed, are truly 2-dimensional, such as wires
in refs. \cite{mo1,mo2,nat}, quantum dots in refs. \cite{mar,piv} etc.
We believe that the observed saturation
can be understood in the frame of our results above, and  
should be due to magnetic scattering.
Including the 
apparent lack of saturation in certain samples - e.g. \cite{nat,pie2}.
In such cases it is difficult to say whether
sufficiently low temperatures have been reached for saturation 
to be observable. The relevant $T$, below which saturation can be observed,
is proportional to the strength of the magnetic scattering rate, and lack 
of saturation was observed in the cleaner samples - e.g. \cite{pie2}.

\vspace{.3cm}

In summary, we have demonstrated that, within the frame of our approach,
dephasing saturation arises from magnetic disorder 
and interactions, with the role of the former being decisive.
Already, without considering interactions \cite{berg}, the cooperon $C_0$ has
a finite correction of the simple diffusion pole - c.f. 
eqs. (\ref{coos}) - which is equivalent to a 'saturating'
dephasing rate, but not for the pure spin-orbit scattering case - c.f.
eqs. (\ref{cooso}). 
We treat the effects of interactions both on the spin-independent
$C_0$ and on the spin-dependent $C_{1,2}$.
Based on symmetry, a lack of saturation is expected for pure 
spin-orbit scattering. 
The magnitude of the 'dephasing
rate' at $T\rightarrow 0$ is set by the magnetic scattering rate.
It appears that our results 
are in agreement with relevant experiments.

\vspace{.3cm}
The author has enjoyed useful discussions/correspondence with N. Birge, 
J. Bird, D.K. Ferry, D.E. Khmelnitskii, 
P. Kopietz, J.J. Lin, P. Mohanty, D. Natelson, 
P. Schwab and J. von Delft.

\vspace{.5cm}
\centerline {\bf APPENDIX A}
\vspace{.4cm}

In the limit $\ep_F \tau \gg 1$ eqs. (\ref{eco1}),(\ref{eco2}) reduce
to:
\bea
C_1 = C_1^o+C_1 ( \Sigma_1 C_1^o+\Sigma_2 C_2^o) +C_2(\Sigma_1 C_2^o
+\Sigma_2 C_1^o) \;\; ,\\
C_2 = C_2^o+C_1 ( \Sigma_2 C_1^o+\Sigma_1 C_2^o) +C_2(\Sigma_1 C_1^o
+\Sigma_2 C_2^o) \;\; .
\eea
Then, setting $u_1=1-(\Sigma_1 C_1^o+\Sigma_2 C_2^o)$ and
$u_2=\Sigma_2 C_1^o+\Sigma_1 C_2^o$,
\be
Det = u_1^2 - u_2^2 \;,\; D_1 = C_1^o u_1+C_2^o u_2 \;, \;
D_2 = C_2^o u_1 + C_1^o u_2 \;\;,
\ee
we have
\be
C_j = D_j/Det \;\;, \;\; j=1,2\;\;.
\ee
Taking 
\be
c = \frac{1}{B+m_1} -\frac{1}{B+m_2 } \;\;, \;
B = Dq^2 -i \om \;\;, \; m_1=2/(3 \tau_{S }) \;\;,\; m_2=2/\tau_{S } \;\;, 
\ee
we obtain
\be
D_j = b_j c + (-1)^j c^2 (b_1^2 - b_2^2) \Sigma_j \;\;.
\ee
Then
\be
C_j = \frac{d_j}{N_+ N_-} \;\;,
\ee
with $d_j = b_j m_0 (B+m_1)(B+m_2)+(-1)^j {m_0}^2 (b_1^2 - b_2^2) \Sigma_j$,
$N_\pm = (B+m_1)(B+m_2)- m_0 (b_1 \Sigma_1 + b_2 \Sigma_2) 
\pm m_0 (b_2\Sigma_1 +b_1 \Sigma_2) $ and $m_0=m_2-m_1$.
We should emphasize that so far this algebra is exact. 

In the following, we only keep terms of order $B$, obtaining :
\bea
d_j=b_j m_0[m_1 m_2+B(m_1+m_2)]+(-1)^j{m_0}^2 (b_1^2 - b_2^2) \Sigma_j\;\;,\\
N_+ = m_1 m_2 +B(m_1+m_2) + m_0 (b_1-b_2)(\Sigma_2-\Sigma_1) \;\;, \\
N_- = m_1 m_2 +B(m_1+m_2) - m_0 (b_1+b_2)(\Sigma_2+\Sigma_1) \;\;.
\eea
Making use of
\be
\frac{1}{N_+ N_-} = \frac{1}{m_0 X} \Big( \frac{1}{N_-} - \frac{1}{N_+} \Big)
\;\;,
\ee
we obtain directly eqs. (\ref{cosog}). 

The derivation above is equally straightforward for the case of 
spin-obit disorder.

\vspace{.5cm}
\centerline {\bf APPENDIX B}
\vspace{.4cm}

In this Appendix we discuss 
the case of finite spin-orbit impurity scattering, i.e.
$\tau_{so}^{-1}>0, \tau_S^{-1}=0$. 
The cooperons are given by \cite{berg,lath}
\bea
C_0^o(q,\om) = \frac{1}{2\pi N_F \tau^2} \;  \label{cooso}
\frac{1}{Dq^2 - i \om + 4/(3 \tau_{so})} \;\;,  \\
C_{1,2}^o(q,\om) = b_{1,2}' \Big\{ \frac{1}{Dq^2 -i \om}
-\frac{1}{Dq^2 -i \om + 4/(3 \tau_{so})} \Big\} \;\;, \nonumber
\eea
with 
$b_1'=(3\tau_{so}/\tau -2)/(2u), b_2'=-1/u, u=4\pi N_F \tau^2$.

Diagonalizing the system formed by eqs. (\ref{eco1},\ref{eco2}) yields for 
the cooperons $C_{1,2}$ 
\be
C_{1,2}(q,\om) = S_{1,2}' \Big[ \frac{1}{Dq^2 -i \om + r_-} - 
\frac{1}{Dq^2 -i \om + r_+} \Big] \;\;. \label{coso}
\ee
Here 
$r_+=(b_1'-b_2')(\Sigma_2-\Sigma_1), r_-=-(b_1'+b_2')(\Sigma_2+\Sigma_1)$,
$S_j'=\{b_j' (Dq^2 -i \om) + M_j'\}/ X', 
X'=2 (b_1' \Sigma_2 + b_2' \Sigma_1)$,
$M_1'=(b_2'^2-b_1'^2) \Sigma_1,\; M_2'=-(b_2'^2-b_1'^2) \Sigma_2$.
The derivation of eqs. (\ref{coso}) is the same as for the magnetic
impurity case given in Appendix A.

The diffusion pole in eqs. (\ref{cooso}) survives, by symmetry, 
in the presence of
the spin-conserving Coulomb interaction, which implies 
$r_+=0$ or $r_-=0$ in  eqs. (\ref{coso}). 

The 2-D weak localization correction to the conductivity is
\bea
\delta \sigma_o = - \frac{e^2 D u}{\pi}
\frac{e H}{\pi} \sum_{n=0}^{N_H} \Big\{
\frac{1}{2\pi N_F \tau^2} \; \frac{1}{4 D e H(n+1/2) + 4/(3\tau_{so})+ 
\tau_{\phi o}^{-1} }   \label{swl2h}  		\\    \nonumber
+ \frac{4 b_2' e D H(n+1/2)+M_2'}{X} \Big( \frac{1}{4 D e H(n+1/2) + r_-}
-\frac{1}{4 D e H(n+1/2) + r_+} \Big)  \Big\}  \;\;,
\eea
with $N_H=1/(4D e \tau H)$. 

Fitting this expression to eq. (\ref{2dwl}) with either $r_+=0$ or $r_-=0$
yields absence of dephasing saturation.
This is the case for all dimensionalities for pure spin-orbit scattering.
As mentioned in the text, this is not the case for finite 
magnetic scattering.

\vspace{.3cm}
$^*$ E-mail address: kast@iesl.forth.gr

\vspace{4cm}

\begin{figure}
\begin{center}
\epsfxsize8cm
\epsfbox{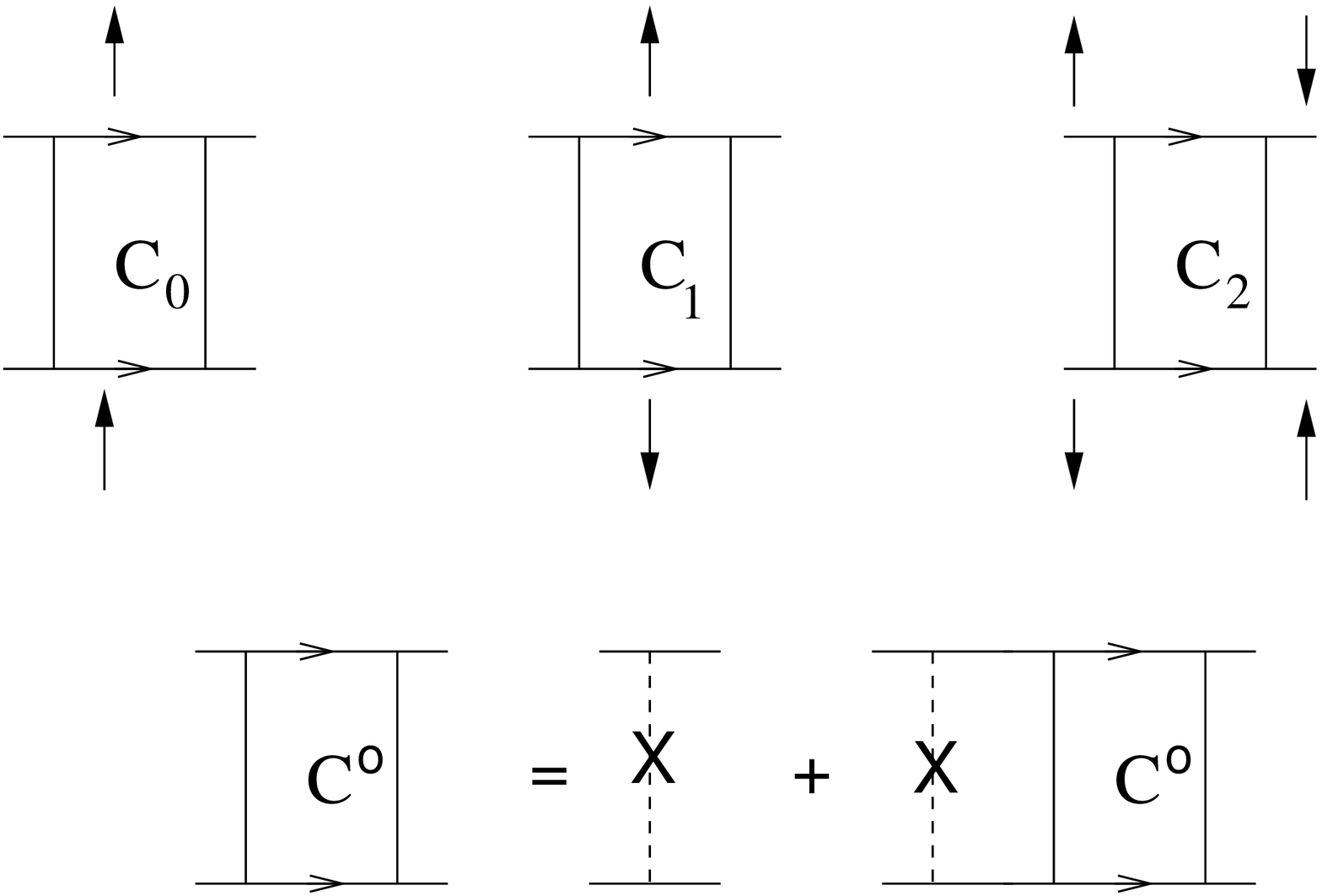}
\vspace{.1cm}
\centerline{Fig. 1}
\end{center}
The 3 cooperons $C_0,C_1,C_2$. Note the spin indices. The cooperons $C_i^o$
do not contain a dephasing rate. The dashed line with the cross stands
for impurity (disorder) scattering. The bare disorder vertices flipping spin,
corresponding to the spin configuration of $C_2$, yield a
coupling of all three cooperons.

\end{figure}

\begin{figure}
\begin{center}
\epsfxsize10cm
\epsfbox{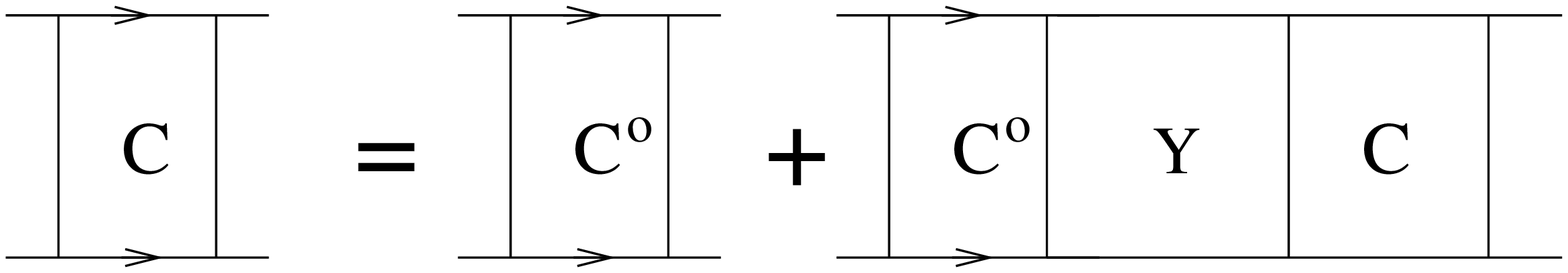}
\vspace{.1cm}
\centerline{Fig. 2}
\end{center}
Schematic form of the equations (\ref{exsc1}-\ref{eco2}) involving the
cooperons and the self-energies $Y$.
\end{figure}

\begin{figure}
\begin{center}
\epsfxsize6cm
\epsfbox{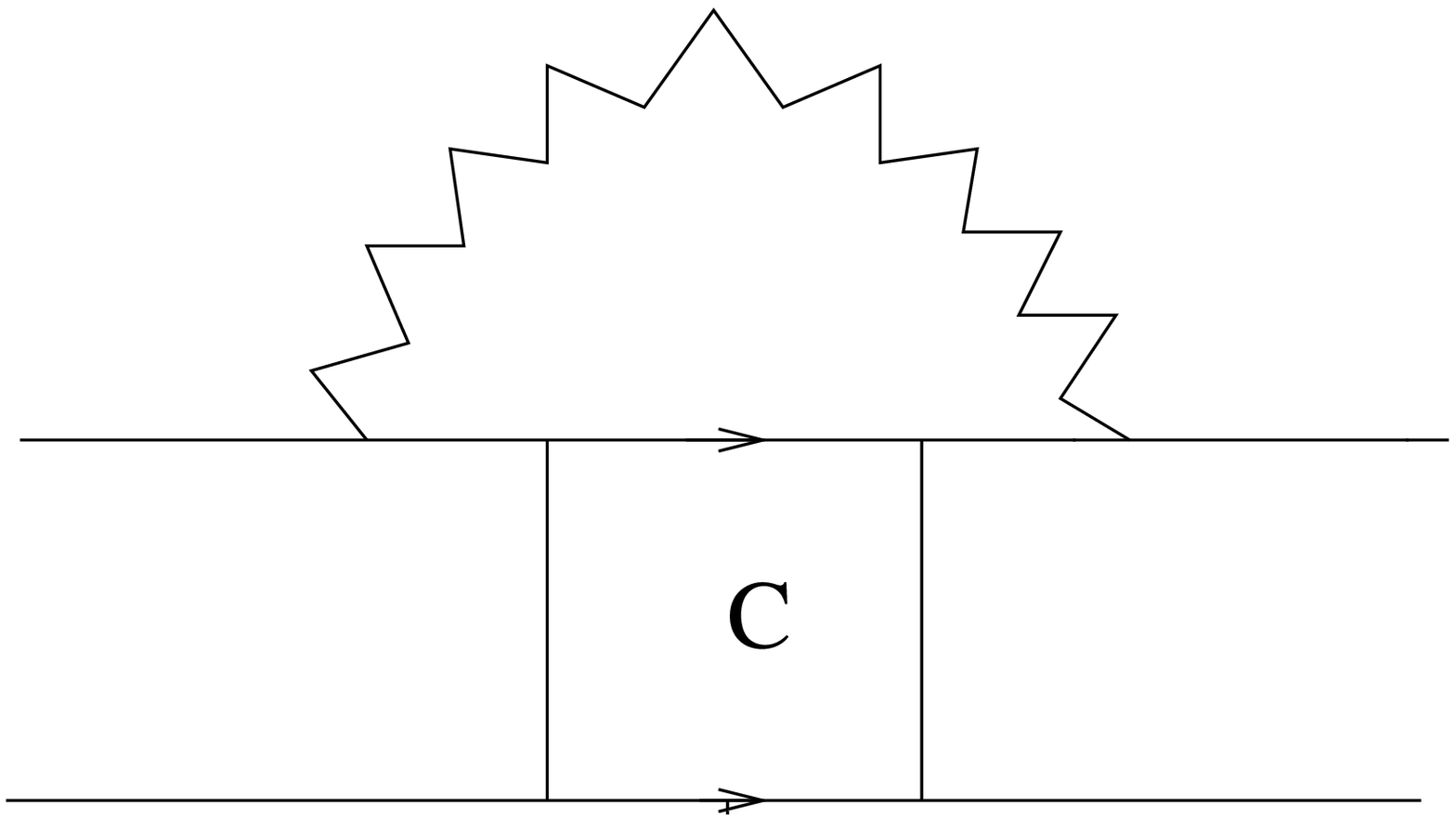}
\epsfxsize6cm
\epsfbox{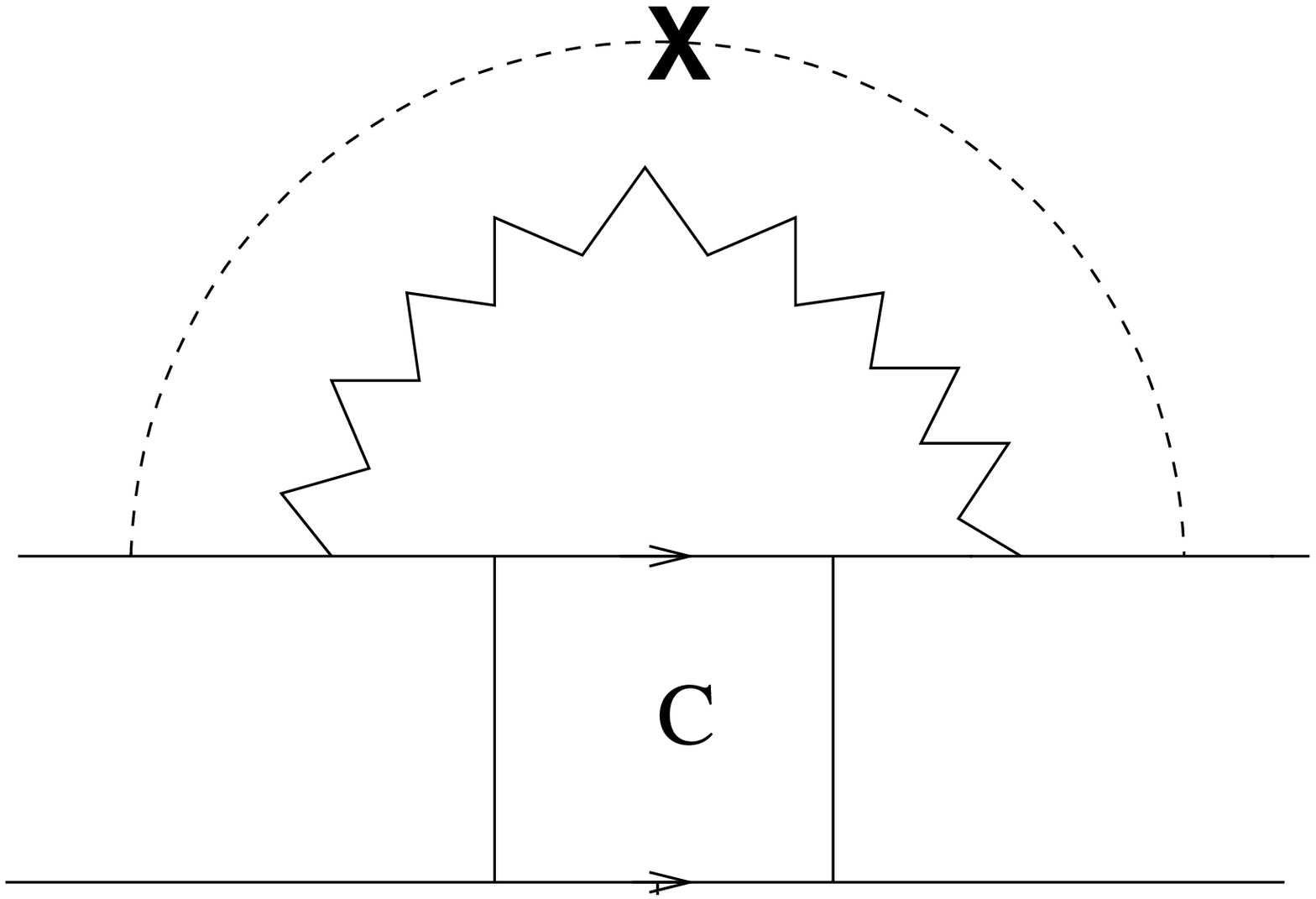}
\vspace{.3cm}
\epsfxsize6cm
\epsfbox{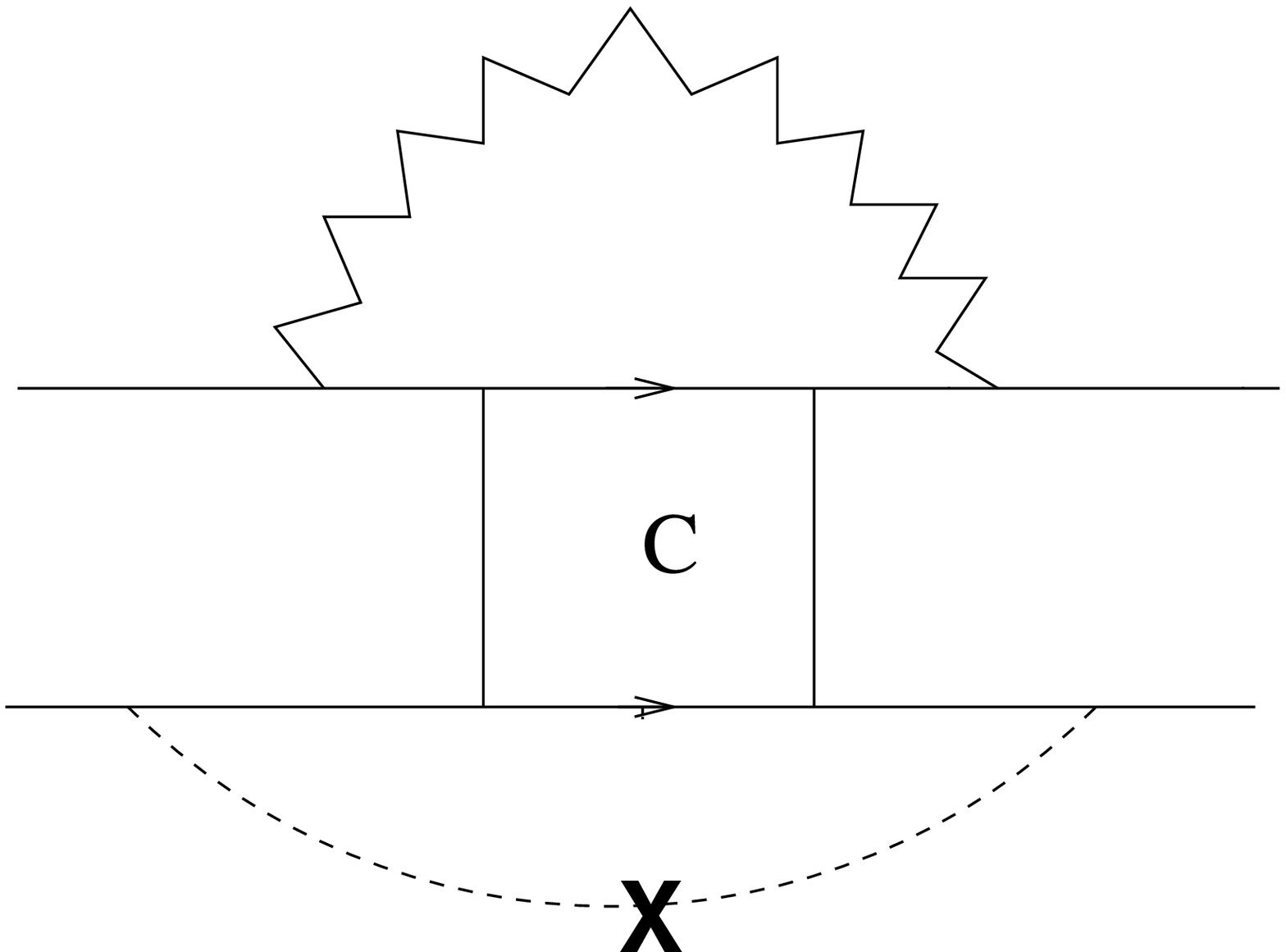}
\vspace{.3cm}
\epsfxsize6cm
\epsfbox{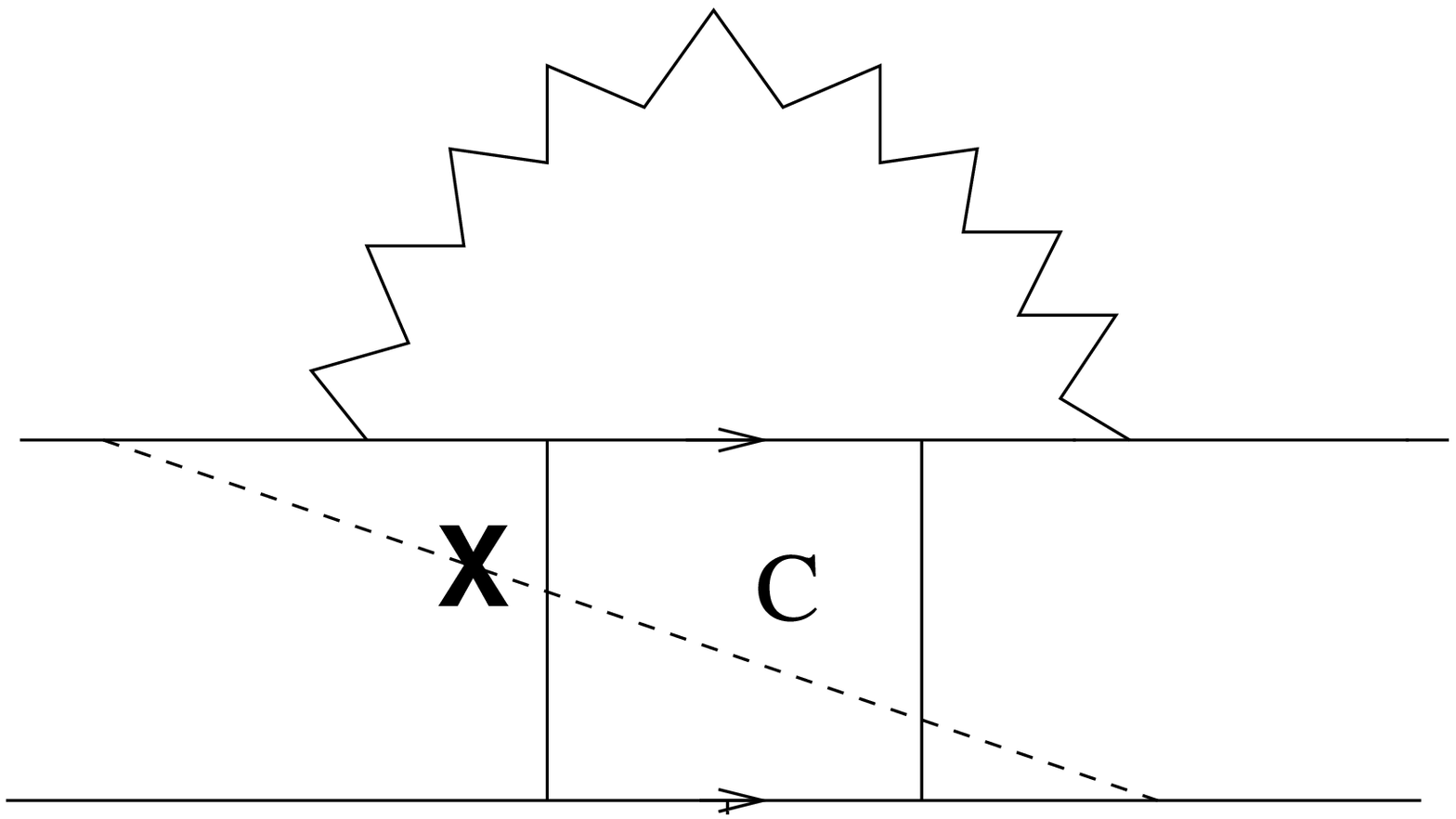}
\centerline{Fig. 3}
\end{center}
The various components of the self-energy $Y$. The wiggly line represents
the screened Coulomb interaction of eq. (\ref{pot}). The diagrams with
the extra spin-disorder impurity line are responsible for the contribution
$d_0 \Sigma_1$ in $\tau_{\phi o}^{-1}$ (c.f. eq. (\ref{ext})). 
We consider {\em all} possible
variations of the diagrams shown here. That is, including terms in which 
the positions of the interaction and impurity vertices are interchanged
along the particle lines. E.g. in the second diagram
above, suppose we label the Coulomb and disorder vertices along 
the upper electron line by the numbers (1,2,3,4). For this diagram we also
consider the permutations (2,1,3,4),(1,2,4,3) and (2,1,4,3).

\end{figure}


\begin{references}
\bibitem{lath}
The prefactors $b_{1,2}$ and  $b_{1,2}'$ in eqs. (\ref{cooso}) and 
(\ref{coos}) are correct herein. In ref. \cite{berg} they contain
a superfluous factor.
\end{references}
\end{document}